# First Detection of the [OIII] 88 μm Line at High Redshifts: Characterizing the Starburst and Narrow Line Regions in Extreme Luminosity Systems

C. Ferkinhoff[1], S. Hailey-Dunsheath[1,2], T. Nikola[1], S.C. Parshley[1], G.J. Stacey[1], D. J. Benford[3], & J. G.Staguhn[3,4,5]

Department of Astronomy

Cornell University

## ABSTRACT

We have made the first detections of the 88 μm [OIII] line from galaxies in the early Universe, detecting the line from the lensed AGN/starburst composite systems APM 08279+5255 at z = 3.911 and SMM J02399-0136 at z = 2.8076. The line is exceptionally bright from both systems, with apparent (lensed) luminosities $\sim 10^{11}$ $L_\odot$. For APM 08279, the [OIII] line flux can be modeled in a star formation paradigm, with the stellar radiation field dominated by stars with effective temperatures, $T_{eff}$ >36,000 K, similar to the starburst found in M82. The model implies ~35% of the total far-IR luminosity of the system is generated by the starburst, with the remainder arising from dust heated by the AGN. The 88 μm line can also be generated in the narrow line region of the AGN if gas densities are around a few 1000 $cm^{-3}$. For SMM J02399 the

[1] Department of Astronomy, Cornell University, Ithaca, NY 14853; cferkinh@astro.cornell.edu
[2] Current Address: Max-Planck-Institut fur extraterrestrische Physik, Postfach 1312, D-85741 Garching, Germany; steve@mpe.mpg.de.
[3] Observational Cosmology Laboratory (Code 665), NASA Goddard Space Flight Center, Greenbelt, MD 20771
[4] Department of Astronomy, University of Maryland, College Park, MD 20742.
[5] Current Address: Department of Physics & Astronomy, Johns Hopkins University, Baltimore, MD 21218.

[OIII] line likely arises from HII regions formed by hot ($T_{eff}$ >40,000 K) young stars in a massive starburst that dominates the far-IR luminosity of the system. The present work demonstrates the utility of the [OIII] line for characterizing starbursts and AGN within galaxies in the early Universe. These are the first detections of this astrophysically important line from galaxies beyond a redshift of 0.05.

*Subject headings:* galaxies: individual (APM 08279+5255, SMM J02399-0136) – galaxies: high-redshift – galaxies: starburst – galaxies: active – submillimeter: galaxies

## 1. INTRODUCTION

The 88 μm [OIII] line is one of eight bright far-infrared fine-structure lines that emerge from electronic ground state configurations of the astrophysically abundant atoms $C^+$ (158 μm), $N^+$ (122 & 205 μm), $N^{++}$ (57 μm), $O^0$ (63 & 146 μm), and $O^{++}$ (88 & 52 μm). These lines are major or even dominant coolants for much of the interstellar medium in dusty star-forming galaxies, and excellent extinction-free probes of the physical conditions of the gas, and/or the strength or hardness of the ambient interstellar radiation fields. Far-IR extragalactic spectral line surveys obtained with the Infrared Space Observatory (ISO) (Malhotra et al. 2001, Negishi et al. 2001, Brauher et al. 2008) detected many of these lines in >200 nearby galaxies. At least 88 galaxies were observed in the [OIII] 88 μm line, with a detection rate near 75%. The [OIII] line was typically the second or third most luminous line (after the [CII] and [OI] 63 μm lines), and in ~10% of the galaxies it was the most luminous far-IR line. The [OIII] line to far-IR continuum luminosity ratio ranges from 0.03 to 2%, with a median value ~0.15%.

To form $O^{++}$ takes 35 eV photons so that hot ($T_{eff}$ > 36,000 K) stars are required for strong [OIII] line emission. Since the [OIII] 88 μm line emitting level is only 164 K above ground, and has a critical density ~510 $cm^{-3}$, the line emission is insensitive to ionized gas temperature, and typically traces moderate density (~100 $cm^{-3}$) clouds. Oxygen is often doubly ionized near AGN, and indeed, the forbidden optical lines of $O^{++}$ (e.g. [OIII] 5007 Å) are among the brightest optical lines from the narrow line regions (NLR) excited by AGN.

We report the first detections of the [OIII] 88 μm line from a galaxy beyond a redshift of ~ 0.05. We strongly (7.0σ) detected the line from the highly lensed (μ~4 to 90, Riechers et al 2009, Egami et al. 2000) broad absorption line (BAL) quasar APM 08279+5255 (hereafter APM 08279) at z = 3.911, and made a weaker (4.1σ) detection of the line from the more moderately lensed (μ ~2.38, Ivison et al. 2010) submillimeter galaxy, SMM J02399-0136 (hereafter SMM J02399) at z=2.8076.

APM 08279 is a composite AGN/starburst system with an *apparent* bolometric luminosity[6] ~7×$10^{15}$ $L_\odot$, of which 2×$10^{14}$ $L_\odot$ emerges in the rest-frame FIR bands (Weiss et al. 2007). Dividing by the largest magnification advocated the luminosity of the system is still ~8×$10^{13}$ $L_\odot$, making it both the system with the largest apparent luminosity known, and an intrinsically extremely luminous system. The strong lensing enables detection and imaging of the source in at least 6 CO rotational transitions, which arise from a disk-like structure with lensed radius ~900 pc containing ~5.3×$10^{11}$/μ $M_\odot$ of molecular gas (Riechers et al. 2009). Through detailed modeling of the far-infrared through radio continuum SED, Riechers et al. (2009) conclude that the dominant heating source for the system is the AGN.

---

[6] We assume a flat cosmology, $\Omega_\Lambda$ = 0.73, $H_0$ = 71 km/sec/Mpc throughout this paper.

SMM J02399 is also a composite AGN/starburst system. It was the first submillimeter galaxy discovered (Ivison et al. 1998), and the first to be detected in a rotational line of CO (Frayer et al. 1998). The molecular gas mass is quite large, ($\sim 2.38 \times 10^{11}/\mu$ $M_\odot$, Ivison et al. 2010), as is the far-IR luminosity ($L_{far-IR} \sim 2.9 \times 10^{13}$ $L_\odot/\mu$, Frayer et al. 1998). The system has at least four distinct components within a 3" diameter region. Initial ground based BVR band imaging revealed two sources: L1, a weak BAL quasar, and L2, a wispy region stretching ~ 3" to the east of L1 (Ivison et al. 1998). Later HST imaging revealed two additional faint, extended (~0.5 to 1") components within the system, labeled L2SW and L1N (Ivison et al. 2010). Based on precise registration between VLA 1.6 GHz, and EVLA CO(1-0) mapping, Ivison et al. 2010 make a strong case that the source of most of the far-IR luminosity is neither L1 nor L2, but rather from L2SW, located roughly between L1 and L2. L2SW has extremely red IRAC colors, and its coincidence with the molecular gas and rest-frame 333 μm continuum emission (centroid adjusted from Genzel et al. 2003) indicates that L2SW is the site of an extreme luminosity, very young starburst that likely dominates the far-IR luminosity of the system.

## 2. OBSERVATIONS

We employed the redshift (z) and Early Universe Spectrometer (ZEUS, Stacey et al. 2004, Hailey-Dunsheath 2009) on the Caltech Submillimeter Observatory (CSO)[7] for our observations. ZEUS is an echelle grating spectrometer that employs a linear 1×32 pixel bolometer array oriented along the dispersion direction that delivers background limited R ~1000 spectroscopy in both the 350 and 450 μm telluric windows. Band-pass filters enable half the array to operate in 5th order (350 μm band) of the echelle, and the other in 4th order (450 μm band) of the echelle so that we obtain simultaneous spectra in the two bands for a single beam on the sky. For APM 08279 (SMM J02399) at redshift of 3.911 (2.8076), the 88.356 μm line is redshifted to 433.916 μm (336.424 μm), and the pixel spacing (~one resolution element) corresponds to 278 (321) km/sec. For both bands, the near diffraction limited ZEUS/CSO beam is ~11" full-width-at-half-maximum, determined by observations of Uranus. At the observed wavelengths, ZEUS achieves sensitivity equivalent to a single-side-band receiver temperature <50 K. Data were taken in standard chop/nod mode with a chop frequency of 2 Hz and amplitude of 30". Spectral response flats were obtained on a cold load, and point-source coupling and calibration (estimated at ±30%) was obtained through observations of Uranus ($T_{Uranus}$ =63 and 71 K at 336 and 434 μm respectively, Hildebrand et al. 1985). On 14 March 2009 we made a 4.5σ detection of the line from APM 08279. This was confirmed at 7σ significance on 1 December 2009 in 77 minutes of on-source integration time with line-of-sight (l.o.s) transmission ~37%. The final spectrum (Figure 1) is this second, high significance integration. We detected (4.1σ) the line from SMM J02399 on 15 January 2010 in 77 minutes of on-source integration time with l.o.s. transmission ~23% (Figure 2).

---

[7] The Caltech Submillimeter Observatory is supported by NSF contract AST-0229008

## 3. RESULTS

### *3.1. Line Luminosity*

Table 1 lists the source line fluxes and apparent luminosities. The line is extremely bright ~$10^{11}/\mu$ $L_\odot$ in both sources. For APM 08279, the [OIII] to far-IR continuum luminosity ratio is ~$5.3 \times 10^{-4}$, and for SMM J02399 it is ~$3.6 \times 10^{-3}$, so that the ratio is ~3 times smaller than the typical value in the ISO surveys for APM 08279, and ~2 times larger for SMM J02399. For both sources, the line is clearly narrow (< 400 to 600 km/sec), so that the line does not arise from broad-line region gas, and therefore either arises from the narrow line region of the AGN or from star formation regions within the system.

With APM 08279, we detected (12σ) the 434 μm continuum at ~348±30% mJy, in very good agreement with the 450 μm SHARC-2 value (342±26 mJy, Beelen et al. 2006) From SMM J02399 the 336 μm continuum was weakly (4σ) detected at ~133±30% mJy. Extrapolating from the 450 μm SCUBA flux (69 mJy) using the dust temperature (45 K) and emissivity index (1.5) typical of luminous IRAS galaxies (Dunne et al. 2000), we predict a 336 μm continuum from SMM J02399 ~98 mJy, in good agreement with our observed value.

### 3.2. Minimum mass of ionized gas.

Without knowledge of the gas density, one may only rigorously calculate a *minimum ionized gas mass* required to support the observed [OIII] line emission. The minimum mass is obtained for the high density, high temperature limit, assuming all the oxygen is in the state $O^{++}$ for the entire HII region. This latter assumption is realistic only for nebula created by very hot ($T_{eff}$ >40,000 K) stars. Under such simplifications, the 88 μm line flux, $F_{10}$ is related to the minimum mass, $M_{min}(H^+)$ by:

$$M_{min}(H^+) = F_{10} \cdot \frac{4\pi \cdot D_L^2}{\frac{g_1}{g_t} A_{10} h\nu_{10}} \frac{m_H}{\chi(O^{++})}$$

where $A_{10}$, and $g_l$, are the Einstein A coefficient, and statistical weight (=3), for the (J=1) 88 μm photon emitting level, $g_t = \Sigma g_i \exp(-\Delta E_i/kT)$ is the partition function, h is Planck's constant, $\nu_{10}$ is 88 μm line frequency, $D_L$ is the source luminosity distance, and $\chi(O^{++})$ is the relative abundance of $O^{++}/H^+$ within the HII region, which, for minimum mass = O/H. Assuming "HII region" gas phase oxygen abundance (O/H = $5.9 \times 10^{-4}$, Savage and Sembach, 2004) for both APM 08279 and SMM J02399 we have $M_{min}(H^+) \sim 3.0 \times 10^9/\mu$ $M_\odot$, ~1% of the apparent molecular gas mass in these systems. For both systems, the minimum ionized gas/molecular mass fraction is small compared with that of the nearby starburst galaxy M82 (~12%, Lord et al. 1996, Wild et al. 1992). Note that the ionized gas mass is inversely proportional to the O/H ratio, and a strong function of the effective temperature of the ionizing stars. If, for example, the HII region is formed by stars with $T_{eff} \sim 36,000$ K, then only ~14% of the oxygen within the HII region is in the form $O^{++}$, so the ionized gas mass is 7 times larger than the value given above.

## 4. DISCUSSION: GAS EXCITATION MECHANISMS

The ionization source for the observed [OIII] line emission can be either UV photons from early type stars, or from the AGN. Both sources are viable for APM 08279, while for SMM 02399 we prefer a star formation model.

### 4.1. *APM 08279*

#### 4.1.1. *The Ionization State of the Gas*.

Very few lines useful for estimating the ionization state of the gas have been detected from APM 08279. However, an upper limit to the 205 μm [NII] line is available (Krips et al. 2007), and this limit can provide a good constraint. Due to their similar ionization potentials, when O is in the form $O^{++}$, N will be in the form $N^{++}$. Therefore, the [NII] 205 μm to [OIII] 88 μm line ratio is very sensitive to the hardness of the local radiation fields. Krips et al (2007) give a 3σ upper limit of 9 Jy-km/sec/beam for the [NII] line within 1000 km/sec of the line center. Integrating over their map, and assuming the [NII] morphology matches the 200 μm continuum morphology, the upper limit scales to ~16 Jy-km/sec, or $1.56\times10^{-19}$ W/m². Therefore the [OIII] 88 μm/[NII] 205 μm line flux ratio is greater than 17.

### *4.1.2. Stars as the Energy Source.*

Rubin (1985) calculates the expected far-IR line intensities for HII regions as a function of gas density, elemental abundances, and stellar effective temperature. We use the "K" model abundances (O/H=$6.76\times10^{-4}$, N/H=$1.15\times10^{-4}$), close to the "HII region" values given by Savage and Sembach (2004), since the dust to gas ratio (reflecting metalicity) appears near the Milky Way value for APM 08279 (Riechers et al. 2009). The lower limit on the [OIII]/[NII] line ratio imposes a strong lower limit of $T_{eff}$ >36,000 K (O9V stars, Vacca et al. 1996). The gas density must be >few cm$^{-3}$ if we assume the ionized gas mass is less than the molecular gas mass, and <10,000 cm$^{-3}$ if we assume that the luminosity of the ionizing stars is less than or equal to the far-IR luminosity of the system. Taking densities between 100 and 1000 cm$^{-3}$ which are typical of starburst regions and AGN-NLRs, and using "K49" models ($10^{49}$ ionizing photons/sec/HII region), the total gas mass traced by the [OIII] 88 μm line is ~$4.3\times10^{10}/\mu$ $M_\odot$ if the HII regions are formed by stars with $T_{eff}$=36,000 K, and 3-8$\times10^9/\mu$ $M_\odot$ if they are formed by hotter ($T_{eff}$=40,000 K) stars. The ionized gas is therefore between 0.6-8% of the molecular gas mass. The total ionization requirement is equivalent to 3-30$\times10^8/\mu$ O9V or 3-10$\times10^7/\mu$ O7.5V stars ($T_{eff}$=36,000, and 40,000 K respectively, Vacca et al. 1996). These ionizing stars generate a total luminosity of 4-40$\times10^{13}/\mu$ $L_\odot$, and 6-20$\times10^{12}/\mu$ $L_\odot$ for the O9V and O7.5V solutions respectively. The lower metalicity (O/H=1.27×10$^{-4}$, N/H =1.47×10$^{-4}$) "D49" models of Rubin (1985) yield the same constraints on ionization, but require about five times the ionized gas mass and three times the number of ionizing stars.

For a starburst model, it is instructive to compare the physical conditions of the starburst in APM 08279 to that occurring in nearby, well studied, and resolved systems such as M82. A comprehensive description of the excitation of the ISM in M82 has emerged from far-IR line studies. The ionized gas component in M82 is well fit by a collection of 60,000 ionization-bounded HII regions, with $n_e$ ~180 cm$^{-3}$, each photo-ionized by a single star generating 10$^{49}$ ionizing photons (Lord et al.1996). The starburst is fueled by a concentration of 8×10$^8$ M$_\odot$ of molecular gas (Wild et al. 1992). Colbert et al. (1999) model the current day stellar population as having evolved from an instantaneous starburst topped by 100 M$_\odot$ stars which occurred 2-3 M yrs ago. This population generates a total stellar luminosity of 3-5×10$^{10}$ L$_\odot$, consistent with the observed M82 far-IR luminosity, L$_{FIR}$ ~2.3×10$^{10}$ L$_\odot$ (Rice et al 1988).

Unfortunately, common star formation tracers like the hydrogen recombination lines and the PAH features, are not useful for a comparison between M82 and APM 08279. The Hα, Paα, and Paβ lines were observed with Spitzer/IRS and AKARI spectrometers, but since the line widths are ~9000 km/sec, and the line fluxes are variable, they clearly arise from the broadline regions of the AGN (Soifer et al. 2004, Oyabu et al. 2009). Relative to the [OIII] 88 μm line, these lines are ~100 times more luminous in APM 08279 than in M82, and so we make no attempt to isolate the narrow component that may be produced in star-forming regions. The Spitzer 6.2 μm PAH feature upper limit (2.3×10$^{-17}$ W/m$^2$, Soifer et al. 2004) is ten times our [OIII] flux. The 6.2 μm PAH/[CII] line ratio is ~3:1 for star forming galaxies (Luhman et al. 2003), and the [CII]/[OIII] ratio increases to ~1:3 for warmer dust galaxies, so that for luminous star forming galaxies one expects a 6.2 μm PAH/[OIII] 88 μm line ratio near unity. Therefore, the Spitzer upper limit does not constrain our models.

An examination of the [OIII] luminosity, far-IR luminosity, and the molecular gas mass of APM 08279 (Table 2) indicates that these tracers are well matched by a superposition of ~3000/μ M82 like starbursts if about 1/3 to 1/2 of the far-IR radiation observed from APM 08279 is attributed to that starburst. The remainder of the far-IR flux would arise from AGN heated dust, which is consistent with the higher dust temperature of APM 08279 (65K, Weiss et al. 2007) as compared to that of M82 (48K, Colbert et al. 1999; 50K, Negishi et al. 2001). If the stellar population is similar to that modeled for M82 (Colbert et al., 1999), then we require the equivalent of $5\times10^8/\mu$ O9V stars and their associated lower mass brethren to heat the dust providing a far-IR luminosity ~$6.9\times10^{13}/\mu$ $L_\odot$. For this case, the HII regions would have gas densities similar to that of M82, $n_e$~180 cm$^{-3}$. The far-IR luminosity due to star formation is equivalent to a star formation rate of 12,000/μ $M_\odot$/year using the Kennicutt (1998) scaling law.

### 4.1.3. The AGN as the Energy Source

It is difficult to quantify the importance of [OIII] line emission from NLR clouds, as these clouds may imitate star forming clouds both in line profile and ionized gas densities. However, by simple scaling arguments we show that emission from these regions of APM 08279 may be important. AGN NLRs have typical gas densities ~100-10,000 cm$^{-3}$, with "average" values near 2000 cm$^{-3}$ (Peterson 1997). In this density range the [OIII] 5007 Å/[OIII] 88 μm line intensity ratio varies from ~0.3 (3) to 65 (180) for "K" ("D") models (Rubin 1985). Observationally, the luminosity of the [OIII] 5007 Å line from AGN is related to the bolometric luminosity of the AGN by $L_{O3}=L_{bol}/3500$, where $L_{O3}$ is the observed luminosity of the [OIII] 5007Å line, uncorrected for extinction (Kauffmann & Heckman, 2005). Since $L_{bol,APM} \sim 7 \times 10^{15}/\mu$ $L_\odot$, the scaling argument suggests $L_{O3} \sim 2 \times 10^{12}/\mu$ $L_\odot$, so that the expected luminosity of the 88 μm line is between $1 \times 10^{10}/\mu$ and $6 \times 10^{12}/\mu$ $L_\odot$, depending on gas density and abundances. The observed 88 μm line luminosity ($1.05 \times 10^{11}/\mu$ $L_\odot$) may arise solely from the NLR of the AGN if this scaling law holds, and if the gas density is ~2000 (1000) cm$^{-3}$, for the "K" ("D") models. The 88 μm line is too weak to support smaller gas density models, and for higher gas densities, only a fraction of the 88 μm line arises from the NLR.

### 4.2. SMM J02399

Ivison et al. 2010 argue that most of the far-IR luminosity in SMM J02399 is due to a massive starburst centered on the source L2SW. Furthermore, since the 6.2 μm PAH flux (~$5 \times 10^{-18}$ W/m$^2$, Lutz et al. 2005), is roughly that of the [OIII] 88 μm line, as expected for star forming galaxies, we construct a model within a starburst paradigm. Ivison et al. (1998) detected the [OIII] 5007 Å line (blended with the [OIII] 4959 Å line) and the Hα line (blended with the [NII] 6548 Å line) from L1 and L2 in the SMM J02399 system. The 5007Å line is exceptionally bright, especially from L2, where within the 1.5" beam, correcting for the expected 3:1 5007Å/4959Å line blending ratio, the 5007Å line flux is $1.5 \times 10^{-17}$ W/m$^2$ ($2.7 \times 10^{11}/\mu$ $L_\odot$). SMM J02399 was not mapped in these lines, but Ivison et al. (2010) mapped the system in Lyα. Assuming the [OIII] and Hα fluxes scale as the Lyα flux we estimate the total [OIII] 5007 Å and Hα line emission from the source is ~2.9 and ~$1.9 \times 10^{-17}$ W/m$^2$ respectively. The [OIII] 5007 Å/Hα ratio (=1.5) is very sensitive to the relative fractions of photons capable of forming O$^{++}$ and H$^+$ (effective stellar temperature), while the [OIII] 5007 Å/88 μm line flux ratio (=4.8) is sensitive to gas density and the electron temperature within the HII region. Good fits to the observed line ratios and absolute fluxes are obtained for $T_{eff}$=40,000 K (O7.5 stars), and gas densities of 100, or 1000 cm$^{-3}$, if one uses the low ("D") or Galactic ("K") metalicities in the Rubin (1985) models respectively. The metalicity and density solutions are degenerate since lower metalicity HII regions have higher electron temperatures, the effect of which is to increase the [OIII] 5007 Å/88 μm line flux ratio at a given density. Correction for extinction effects on the optical lines increases their flux, raising the requisite gas density by factors roughly equal to the extinction correction factor. The requisite ionized gas mass is ~3.3-30×10$^9$/μ M$_\odot$, where the lower value is for "K" metalicities. For both cases, the ionizing flux is equivalent to that of 1.0×10$^8$/μ O7.5 stars. The absorption of the full luminosity of these stars by dust would account for ~75% of the observed far-IR luminosity of the system. The inferred star formation rate is ~5000/μ M$_\odot$/year using the Kennicutt (1998) scaling law.

## 5. SUMMARY AND OUTLOOK

We have made the first detections of the [OIII] 88 μm line from galaxies in the early Universe. The line can trace either young stars or the NLR near AGN. The observed 88 μm line emission from APM 08279 can arise from a starburst whose hottest stars have $T_{eff}$ >36,000 K, or from the NLR of the AGN. From SMM J02399 the 88 μm line is modeled as arising from a massive starburst dominated by very early type ($T_{eff}$ >40,000 K) stars most likely located near the extended source, L2SW. To further constrain the physical properties of the emitting gas requires additional extinction-free spectral line probes. For example, gas density is constrained with the detection of the [OIII] 52 μm line. The density-sensitive 52 μm/88 μm [OIII] line intensity ratio is ~0.7 for low density clouds rapidly growing to the saturation value of 10 at densities above 10,000 cm$^{-3}$. Unfortunately, for both sources, the 52 μm lines are redshifted into regions of the far-IR/submm bands that are totally blocked by Earth's atmosphere. However, the expected line fluxes (~2 to 60×10$^{-18}$ W/m$^2$) are within reach of the Herschel SPIRE spectrometer in 1 to 40 hours of integration time, so that Herschel SPIRE observations should provide important density constraints on our models. It is also likely that the [OIV] 26 μm line will be detectable from these systems with the Herschel PACS spectrometer. The [OIV]/[OIII] line ratio is a sensitive near extinction-free indicator of the hardness of the ambient radiations fields, and will strongly distinguish a starburst from AGN origin for the extreme far-IR luminosity from these two systems.

## ACKNOWLEDGEMENTS

This work was supported by NSF grants AST-0096881, AST-0352855, AST-0705256, and AST-0722220, and by NASA grants NGT5-50470 and NNG05GK70H. We thank our colleagues T.L. Herter and H. Spoon for useful discussion, and the CSO staff for their excellent support of ZEUS operations. We also thank the reviewer for his prompt and helpful comments on this paper.

| Table 1: Source Parameters | | | | | | | |
|---|---|---|---|---|---|---|---|
| Source | RA(J2000) | Dec(J2000) | z | $D_L$(Gpc) | $L_{far-IR}$ ($L_\odot$) | F([OIII]) ($10^{-18}$ W/m$^2$)[1] | L([OIII]) ($L_\odot$) |
| SMM J02399 | $02^h39^m51.9^s$ | -01°35'59" | 2.8076 | 23.8 | 2.9E13[2] | 6.04±1.46 | 1.06E11 |
| APM 08279 | $08^h31^m41.6^s$ | 52°45'17" | 3.911 | 35.6 | 2.0E14[3] | 2.68±0.38 | 1.05E11 |

[1]Statistical uncertainties only. Calibration uncertainties are ~ 30%. [2]Weiss et al. 2007, [3]Frayer et al. 1998.

| Table 2: Comparison Between APM 08279 and M82 | | | | | | |
|---|---|---|---|---|---|---|
| | [OIII] 88 µm Luminosity ($L_\odot$) | H α Luminosity ($L_\odot$) | Far-IR Luminosity ($L_\odot$) | Molecular Gas Mass ($M_\odot$) | [OIII]/Far-IR | [OIII] 88 µm/ [NII] 205 µm |
| APM 08279 | 1.05E11[2] | 1.26E13[4] | 2.0E14[6] | 5.3E11[8] | 5.3E-4 | > 17 |
| M82 | 3.5E7[1] | 1.8E7[3] | 2.3 – 3.2 E10[5,1] | 1.8E8[7] | 1.1 – 1.5E-3 | 15.5[9] |
| APM/M82 ratio | 3060 | 682000 | 6300 – 8700 | 3000 | 0.35 – 0.48 | |

Notes: [1]Colbert et al. 1999, [2]Present work, [3]McLeod et al. 1993, [4]Oyabu et al. 2009, [5]Rice et al. 1988, [6]Beelen et al. 2006, Weiss et al. 2007, [7]Wild et al. 1992, [8]Riechers et al. 2009, [9]Lord et al. 1996, Watson et al. 1984

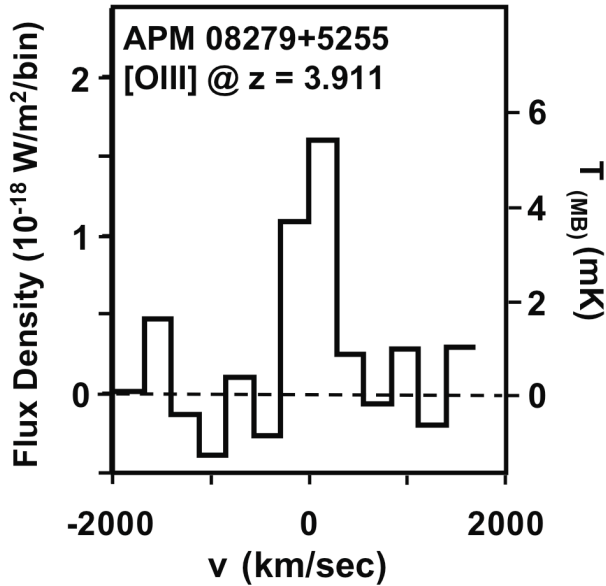

**Fig. 1.** ZEUS/CSO detection of the [OIII] 88 µm line from APM 08279+5255. Velocity is referenced to z=3.911. The continuum emission has been subtracted off.

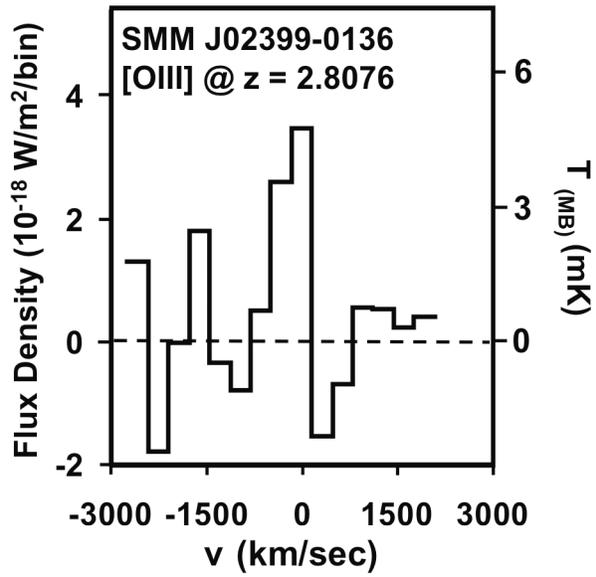

**Fig. 2.** ZEUS/CSO detection of the [OIII] 88 μm line from SMM J02399-0136. Velocity is referenced to z=2.8076. The continuum emission has been subtracted off.